# FIELD DESORPTION ION SOURCE DEVELOPMENT FOR NEUTRON GENERATORS

I. Solano, B. Reichenbach, P. R. Schwoebel, D. L. Chichester, C. E. Holland, K. L. Hertz, and J. P. Brainard.

A new approach to deuterium ion sources for deuterium-tritium neutron generators is being developed. The source is based upon the field desorption of deuterium from the surfaces of metal tips.

Field desorption studies of microfabricated field emitter tip arrays have been conducted for the first time. Maximum fields of 3 V/Å have been applied to the array tip surfaces to date, although achieving fields of 2 V/Å to possibly 2.5 V/Å is more typical. Both the desorption of atomic deuterium ions and the gas phase field ionization of molecular deuterium has been observed at fields of roughly 2 V/Å and 2-3 V/Å, respectively, at room temperature. The desorption of common surface adsorbates, such as hydrogen, carbon, water, and carbon monoxide is observed at fields exceeding ~1 V/Å. *In vacuo* heating of the arrays to temperatures of the order of 800˚C can be effective in removing many of the surface contaminants observed.

## I. INTRODUCTION

Several National and Homeland Security activities involve the detection and identification of special nuclear material (SNM) including highly enriched uranium and plutonium. While passive detection systems can be used in some cases, for the most challenging problems involving shielded SNM it is generally accepted that these materials can only be reliably detected using active interrogation techniques such as neutron interrogation. However, before large-scale neutron interrogation systems can be reliably deployed for this purpose improvements are needed to enhance the field performance attributes of the related ion source and detector technologies; in particular, the National Academy of Sciences has concluded that significant advances in neutron generator technology are required [1].

Existing compact, sealed-tube neutron generators create and then accelerate deuterium (*D*) and/or tritium (*T*) ions into *D* and/or *T*–hydrided metal targets to produce neutrons by the *D-D* or *D-T* nuclear fusion reactions $D + D \rightarrow n + {}^3He$ and $D + T \rightarrow n + {}^4He$. With few exceptions the ion sources in these generators extract ions from a plasma created in various types of electrical discharges [2]. This leads to generators with either long lifetimes (1000 h) and low outputs (~$10^8$ n/s), such as well logging generators, or high outputs (>$10^{12}$ n/s) and short lifetimes (a few hours). For field detection applications one would like to have long-lived, high output generators. Additional desirable generator features include variable pulsing ability, rugged design, low cost, ambient cooling and, in the case of man portable systems, high energy efficiency.

To improve the performance of existing neutron generators we are developing a novel approach for the ion source [3,4]. This ion source uses the electrostatic field desorption (EFD) [5] of deuterium and/or tritium from the surface of metal tips to create atomic ions. The yield of a neutron generator scales with the ion beam current which in turn, for the EFD source, scales with the tip surface area and number of tips. Thus, as shown schematically in Figure 1, an array of tips is being used for the EFD source.

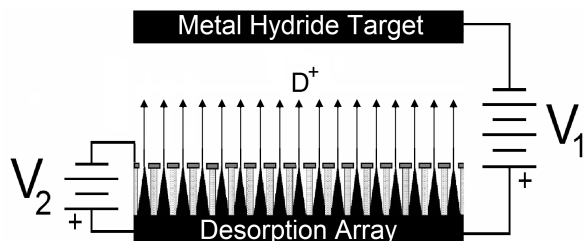

*FIG 1: Schematic diagram of the deuterium-tritium neutron generator using a field desorption deuterium ion source. Voltages V1 and V2 control the deuterium ion accelerating voltage and current respectively.*

The EFD source operates in a pulsed mode wherein a voltage pulse ($V_2$ in Figure 1) applies an electric field sufficient to remove, as ions, deuterium and/or tritium atoms that have adsorbed on the tip surfaces from the gas phase. The hydrogen isotope gases are subsequently reabsorbed, at a rate dependent upon the background gas pressure, before another pulse is applied.

Potential key benefits of this type of ion source relative to conventional ion sources are: 1) increased electrical efficiency and reduced power requirements relative to neutron yield to enable effective man-portable systems, 2) scalable neutron output (with the dimensions of the desorption array) to meet different application needs, 3) increased generator and target lifetime due to the use of a distributed ion beam, and 4) a reduction in system complexity due to the elimination of plasma based ion sources.

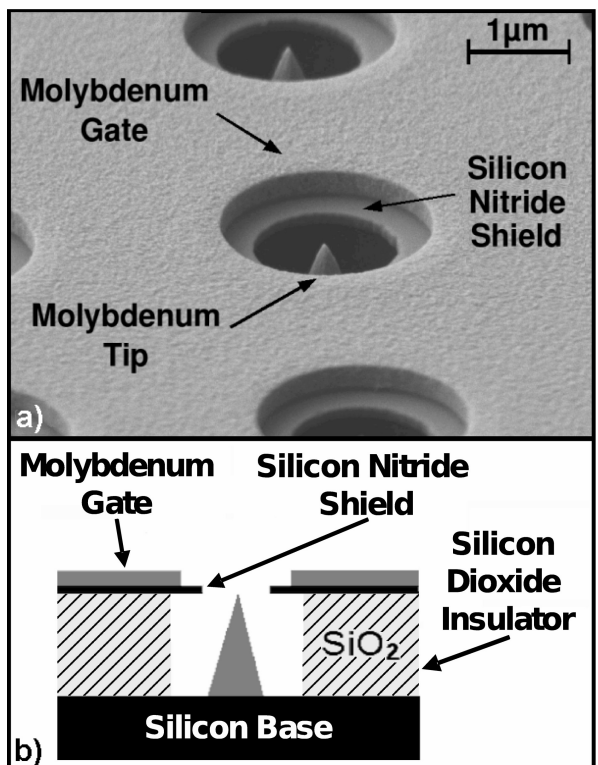

*FIG 2: The modification performed to standard Spindt-type arrays to increase the operating voltage. (a) Scanning electron micrograph of a $Si_3N_4$ shielded array. (b) Schematic of the Spindt-type array with a shield of $Si_3N_4$ dielectric.*

Here we report the first studies using modified Spindt-type tip arrays for an EFD ion source, see Figure 2. The integral gate electrode of these microfabricated tip array assemblies permits a high tip packing density, and thereby the potential for large ion currents. This electrode also allows for the straightforward modulation of $V_2$ using relatively low voltages (~1 kV) to achieve field desorption.

## EXPERIMENTAL

The experimental chamber is an ultrahigh vacuum imaging atom probe [6] achieving a base pressure in the low $10^{-10}$ Torr range following bakeout at ~ 200˚C for 12 h. Species desorbed from the tip surfaces are identified using time-of-flight (TOF) mass analysis with typically an ~12 cm drift distance, and detected using a chevron microchannel plate employing a *P-47* phosphor screen. Referring to Figure 1, in the atom probe the target is replaced with the chevron detector and $V_1$ is set to zero in order to obtain a field free drift region between the gate (i.e. grid) electrode and the detector. An Amperex XP2262 photomultiplier tube monitors the output of the phosphor screen and the characteristic ~60 ns phosphor decay time is thus visible in the time-of-flight mass spectra. The voltage pulse applied to the tip electrode ($V_2$ in Figure 1) of the tip arrays to initiate ion desorption was (at the pulse generator) a square wave with a duration of either 20 ns or 100 ns with a rise time of ~2 ns into a 50 Ω load. Analysis of the field electron emission characteristics of the arrays when a negative going pulse was applied to the tip electrode showed a rise time of ~ 4 ns due to impedance mismatches involved in transmitting the pulse to the array. For these experiments only hydrogen or deuterium gas loadings were used in the chamber, tritium was not used.

The tip arrays were 50 000-tip Spindt-type field emitter arrays having molybdenum tips with gate-to-base oxide thicknesses of either 0.5 μ m or 1.75 μ m [7]. The short tip-to-gate hole rim distance (typically <1 μ m) means that mass identification can be difficult because the picosecond flight times of the ions in the acceleration gap are much less than the rise-time of the voltage pulse used to initiate desorption. This allows for variation in the final energy of the ions and therefore greater uncertainty in the ion mass-to-charge ratio (*au/e*). The TOF system could be calibrated for mass-to-charge ratios less than or equal to 4 *au/e* by filling the system with $H_2$ and/or $D_2$ gas and observing the arrival of, for example, the field ionized molecular ion or atomic fragment. Ion species exceeding 4 *au/e* are estimated from this low mass calibration procedure. Unless noted otherwise, the tip array field–voltage proportionality factor used to estimate ion emission fields was calculated assuming the onset of electron emission from the arrays (~50 nA) for a given voltage occurred at an electric field at the tip apices, *F*, of 0.25 V/Å. *In vacuo* heating of the arrays was accomplished using a custom alumina insulated filament to which the TO-5 header holding the arrays was attached. Array temperatures were measured using either an iron-constantan thermocouple attached directly to the TO-5 header or a disappearing filament optical pyrometer. Research grade (99.9999%) hydrogen and/or deuterium gas was admitted directly into the atom probe chamber from 1- l glass flasks. Studies were conducted at room temperature (293 K) and at 77 K. Results from the studies done at 77 K will be presented in a future paper [8].

## RESULTS

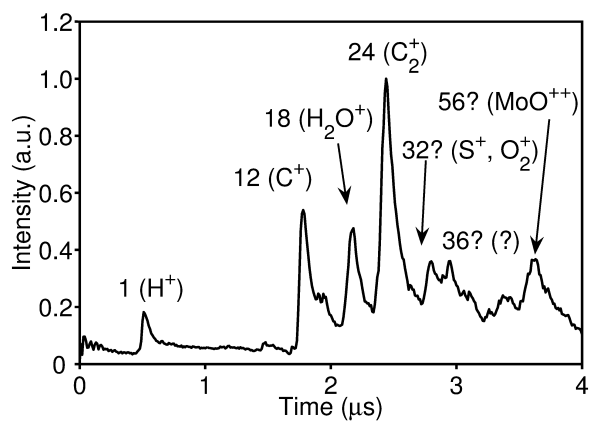

*FIG 3: Common surface contaminants removed by field desorption from as-fabricated arrays. (V = 410 volts, F = 1.7 V/Å).*

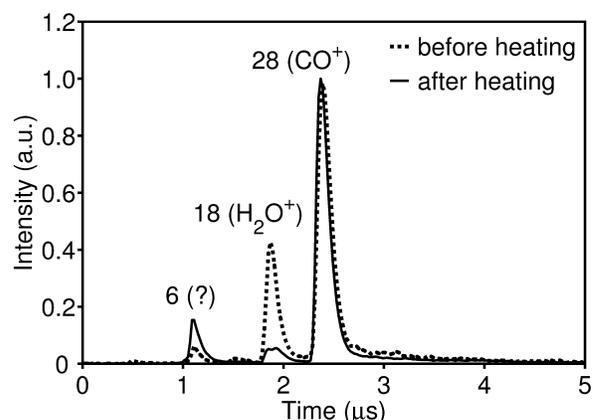

*FIG 4: The removal of $H_2O$ by heating. Time of flight spectra of an as-fabricated array and the array after heating in situ to 800°C for 40 min. (V = 425 V, F = 1.9 V/Å).*

Ion emission was first attempted using standard Spindt-type field emitter arrays (0.5 μm thick oxides) at $F < 0.7$ V/Å (i.e. typically voltages of $V_2 <$ 250 V). Attempting to reach fields needed for desorption of atomic hydrogen or deuterium ($F > 1$ V/Å) led to electrical breakdown and damage to the array structures. Two array structure modifications were made that increased the fields achievable at the tip surfaces. The first was to increase the thickness of the insulating oxide layer between the gate and base electrodes from 0.5 μm to 1.75 μm. The second was to incorporate an insulating shield of $Si_3N_4$ in order to suppress electron emission from the gate, see Figure 2. This type of array was used for the remainder of the studies reported herein.

Atom probe studies of as-fabricated arrays, i.e. those that had undergone no post fabrication treatment except hydrogen firing and storage in vacuum until use [9], typically showed a wide variety of surface adsorbates. Figure 3 is an example of a mass spectrum showing some of the commonly observed species such as, mass-to-charge ratios of 1 ($H^+$), 12 ($C^+$), 18 ($H_2O^+$), 24 ($C_2^+$), and 32 ($S^+$, $O_2^+$). Other commonly observed species were 16 ($O^+$) and 28 ($CO^+$). The presence of compounds containing $H$, $C$, $O$, and $S$ on metal surfaces exposed to the atmosphere and subjected to only a mild vacuum bakeout (~ 200°C), are not unexpected [10].

Adsorbate removal by heating in vacuum could yield predictable results, see Figure 4. In this example, the initial desorption spectra, taken following installation in vacuum and system bakeout, showed the presence of $H_2O$ (18) and $CO$ (28). After heating in vacuum at ~860°C for 40 min. the level of $H_2O$ has decreased significantly while the level of $CO$ has remained essentially unchanged. This is consistent with the fact that $CO$ is only totally removed from Mo at temperatures above ~1400°C [11,12]. Typically, longer heating times (>10 h) at temperatures of ~800°C are required to clean the tip surface sufficiently to allow for the adsorption of deuterium.

Future work in this area will permit the use of more advanced cleaning processes including higher temperature hydrogen and vacuum firing. These techniques are commonly employed in the manufacture of sealed tube neutron generators.

Figure 5 is a spectrum from an as-fabricated array in an atmosphere of $10^{-5}$ Torr $D_2$, at an applied field of 1.6 V/Å where $H$, $H_2O$ and $C$ are observed. The source of $H$ is as yet unclear. At times spectra containing only $H^+$ are observed from the as-fabricated arrays suggesting desorption of hydrogen adsorbed from the residual gas atmosphere of the vacuum. When observed with the carbon, dissociation of some hydrocarbon species cannot be ruled out. Additional experiments are being conducted to determine the source of hydrogen in various circumstances. Figure 6 shows the TOF spectrum from the same array in $10^{-5}$ Torr of $D_2$ after heating in vacuum for 12 h at ~780°C. A significant atomic deuterium peak is now observed. Apparently the level of surface contamination present on the tips of the as-fabricated array inhibited the adsorption of deuterium [12]. This is not surprising, as adsorbate replacement processes are governed by sublimation energetics and the binding energy of hydrogen is lower than species such as $H_2O$ and $CO$, for example.

To confirm the deuterium peak was due to desorption and not gas phase field ionization of deuterium, the gas was evacuated to $10^{-9}$ Torr and additional spectra accumulated, as shown in Figure 6. Here there is only an initial slight decrease in the height of the atomic deuterium peak, consistent with desorption of deuterium from the tip surfaces. If the majority of the signal was due to gas phase ionization the peak height would decrease dramatically as the result of such a large decrease in deuterium partial pressure.

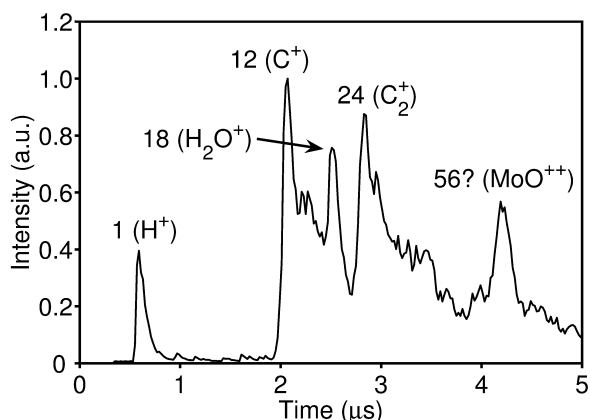

FIG 5: Time of flight spectra of surface contaminants on an as-fabricated array in a partial pressure of 10-5 Torr of deuterium gas. (V = 235 V, F = 1.3 V/Å).

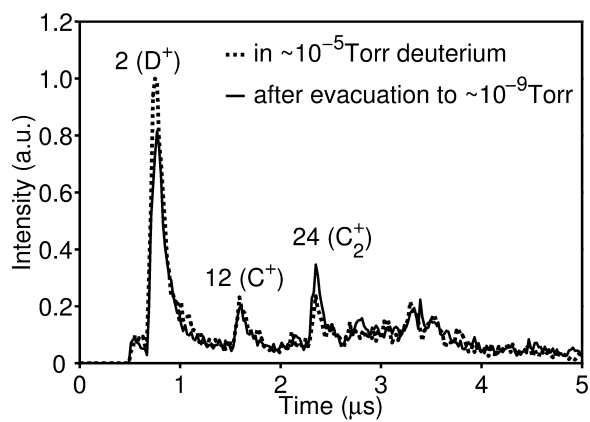

FIG 6: Observation of field desorption of deuterium. Time of flight spectra of the array used in Figure 5 after heating to 860°C for 12 hours in vacuum. Initial spectra taken in $10^{-5}$ Torr of deuterium gas, and subsequent spectra taken after evacuation of the deuterium gas.

According to field desorption studies of deuterium from single wire tips [13], the desorption of deuterium should occur at fields on the order of 3 V/Å. In the case of the arrays, desorption is observed at fields of ~ 2 V/Å. The reduced field required for desorption could be due to the presence of residual surface contamination onto which the deuterium is adsorbed. As we have only heated the arrays to temperatures of ~800°C and not yet achieved fields sufficient to field evaporate the tip surfaces (~ 4 V/Å for molybdenum) the presence of residual surface contamination would not be surprising.

Following the accumulation of the spectra associated with Figure 5 and 6, the array was allowed to remain in UHV for approximately 5 days. After this waiting period desorption spectra at ~2 V/Å and $10^{-5}$ Torr $D_2$ were then dominated by the presence of CO, see Figure 7. This is not surprising as CO typically has a significant partial pressure in baked stainless-steel chambers [14], and would accumulate on the surface at the expense of hydrogen (or its isotopes) [12]. Note that these data also confirm that the deuterium signals shown in Figures 5 and 6 are due to desorption and not field ionization, as deuterium signals were not observed under similar applied fields and gas pressures.

It is clear that surface contaminants can inhibit the adsorption of deuterium onto the tip surfaces, a phenomenon that is also in agreement with our studies on single etched-wire molybdenum tips [13] and such processes generally. Such contaminants can reach the high field region of the tip surfaces either by adsorption from the gas phase or by diffusion from the tip shank to the high field region. In the case of CO adsorption shown in Figure 7, the time required to accumulate the material is consistent with adsorption from the gas phase, here 150 h at $10^{-10}$ Torr. We assume a standard partial pressure of ~ $10^{-11}$ Torr of CO and an initial sticking probability of ~1 [15], meaning that ~100 hrs is required to form a monolayer of CO.

There are cases where surface diffusion is the major cause of contaminants appearing in the high field region. In this process, contaminants are first removed by field desorption. The resulting contaminant concentration gradient drives surface diffusion to repopulate the high field region with contaminants present on the shank region of the tips. Note that contaminants can also be drawn to high field regions by the field gradient (i.e. polarization forces); however, unlike in conventional atom probe experiments this does not appear to be a major factor here because only pulsed fields as opposed to pulsed and d.c. bias fields are used.

Figure 8, shows an inverse relationship between the pulse rate and the amount of contaminants desorbed from the tip apices. The spectrum was taken at an operating pressure of 1x$10^{-9}$ Torr (consisting of principally $D_2$) and the pulse rate varied from 0.1 to 10 seconds delay between pulses. With monolayer formation rates due to adsorption from the gas phase at a maximum of ~$10^4$ s /monolayer (for $D_2$ with a sticking probability of ~ 0.1 on clean molybdenum) no significant adsorption should occur on time scales of 0.1 to 10 s. Thus it is

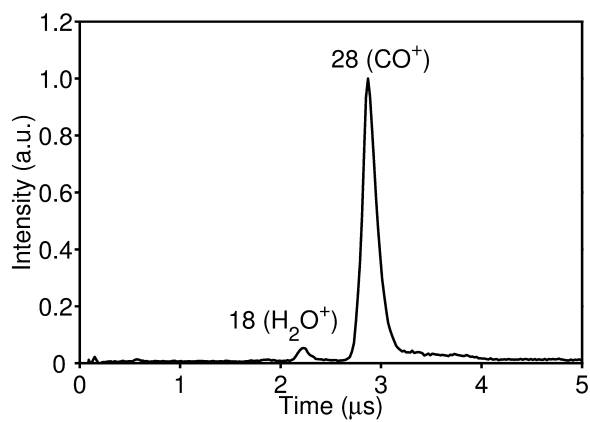

FIG 7: *The accumulation of CO in the high field region of the tip surfaces with time. Time of flight spectrum of the array used in Figure 5 and 6 operating in $10^{-5}$ Torr of deuterium gas after being at $10^{-10}$ Torr for 150 hours. (V = 335 V, F = 1.9 V/Å)*

reasonable that the observed changes in contaminant levels shown in Figure 8 are due to surface diffusion. Similar experiments[8] conducted at 77 K show a decreased contamination rate, also consistent with surface diffusion processes. Note that the trend observed in Figure 8 also rules out the possibility that a significant fraction of the species observed were produced by gas phase ionization. If there was a significant component due to gas phase ionization, the intensity level of the ions observed would not vary with the pulsing rate.

Gas phase field ionization has been observed at room temperature at fields exceeding ~2 V/Å, in partial pressures of deuterium. Figure 9 is a TOF spectrum at a field ~ 2 V/Å and 3 V/Å in a $D_2$ partial pressure of $10^{-5}$ Torr. As the field is increased the ratio of atomic to molecular deuterium increases due to field dissociation of the parent molecule [16]. If the applied field was not limited by voltage breakdown of the microfabricated structure and could be increased further, eventually only atomic deuterium would be observed. The fields predicted using the onset of electron emission are seen to be in reasonable agreement with the fields expected for the onset of hydrogen ionization and dissociation.

## CONCLUSION

Atom probe studies of field emitter arrays have been conducted for the first time. To date, pulsed voltages applied to the microfabricated tip arrays have been typically limited to ~550 V (~2.5 V/Å maximum) due to electrical breakdown. It remains unclear as to whether this limit is related to vacuum, surface, or bulk breakdown processes. We are presently fabricating tip structures with roughly 5 µm of insulating oxide in order to increase operating voltages.

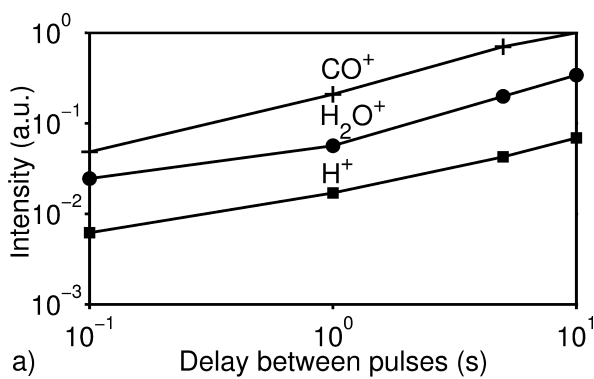

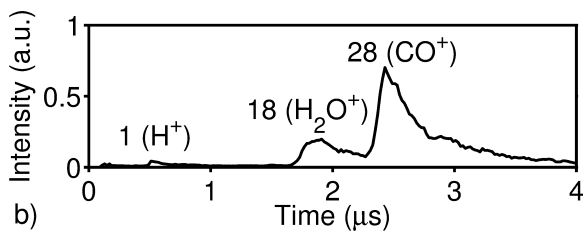

FIG 8: *Surface diffusion of adsorbates into the high field region of the tip surfaces. TOF spectra show (a) the variation in the quantity of the species observed as a function of delay time between pulses and (b) the spectrum at a delay time of 5 seconds.*

The desorption of atomic deuterium at room temperature from modified microfabricated Spindt type arrays has been observed. Atomic deuterium is often accompanied by the desorption of other mass species involving compounds of carbon, oxygen and hydrogen. Absolute measurement of the number of ions desorbed in a single pulse is presently being quantified.

Surface contamination caused by either diffusion along the shank of the tips or adsorption from the gas phase can have a significant impact on the spectra observed. Cleaning procedures such as field desorption and heating in vacuum to temperatures of the order of 800°C significantly reduce the observed levels of contamination. Surface cleaning procedures involving thermal annealing, hydrogen plasma treatment and laser ablation are being investigated as a means to enhance the percentage of deuterium ions relative to all others desorbed from the tip. Undoubtedly, field evaporation of the tips themselves will ultimately be useful for cleaning the tip surfaces and enhancing the uniformity of the ion emission from tip to tip in the array. Contaminant species identification could be made more accurate using pulsed laser desorption, as this circumvents the potential problem of having short flight times in the acceleration gap with voltage pulsing [6].

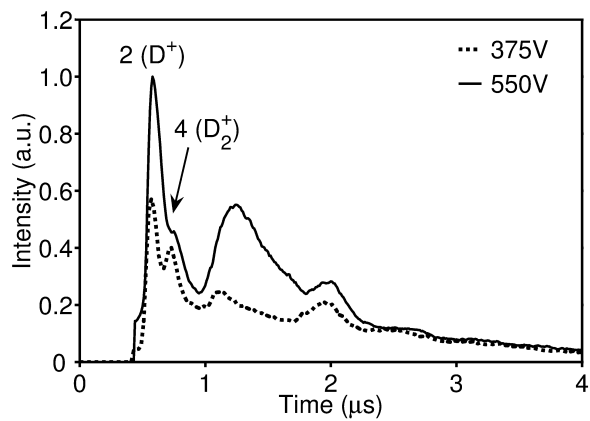

*FIG 9: The observation of gas phase ionization of deuterium. TOF spectrum of an array in $10^{-5}$ Torr of deuterium gas at V = 375 V, F = 2.1 V/Å and at V = 550 V, F = 3.1 V/Å.*

## ACKNOWLEDGEMENTS

The fabrication of vacuum hardware by John DeMoss and the electronics hardware by John Behrendt of the UNM Physics and Astronomy Department shops is gratefully acknowledged. This work was supported by the U.S. Department of Energy through the National Nuclear Security Administration's Office of Non-proliferation Research and Development (NA-22).